\documentclass{emulateapj}
\usepackage{subfigure}

\shorttitle{The softening phenomenon} \shortauthors{Qin}

\begin{document}

\title{Roles of the kinetic and dynamic mechanisms in the $L_p - E_p$ relation}

\author{Yi-Ping Qin\altaffilmark{1,2,3}, Zhi-Fu Chen\altaffilmark{2,1}}
\altaffiltext{1}{Center for Astrophysics, Guangzhou University,
Guangzhou 510006, P. R. China; ypqin@gzhu.edu.cn}
\altaffiltext{2}{Department of Physics and Telecommunication
Engineering, Baise University, Baise, Guangxi 533000, P. R. China}
\altaffiltext{3}{Physics Department, Guangxi University, Nanning
530004, P. R. China}

\begin{abstract}
The $L_p - E_p$ relation is a well-known relation in gamma-ray
bursts. Its implication remains unclear. We propose to investigate
the underlying mechanisms of this relation by considering the
corresponding kinetic and dynamic mechanisms separately. In this
way, one can tell how much the kinetic or dynamic mechanism
contributes to the index of the relationship. Our analysis gives
rise to several conclusions. (1) The index of the kinetic effect in
the $L_p - E_p$ relation can simply be derived from the theory of
special relativity, which is generally larger than 2, depending on
the situation concerned. (2) The index of the dynamic effect in the
relation can be deduced from observation once a model of jets is
adopted. According to current GRB data, we find: the dynamic effect
alone tends to make an anti-correlation between $L_p$ and $E_p$; in
terms of statistics, the dynamic effect is obviously smaller than
the kinetic effect; in the situation of jets with moving discrete
radio clouds which moves directly towards the observer, the index of
the dynamic effect is currently constrained within $(-1.6, -1)$,
while in other situations of jets, the constrains are different;
both internal and external shocks can account for the current data.
\end{abstract}


\section{Introduction}
Gamma-ray bursts (GRBs) are among the most violent astrophysical
events ever observed in the universe. The objects are detected
mainly at gamma-ray bands (Klebesadel et al. 1973) and many of them
are followed by afterglows which can be detected at X-ray or lower
energy bands (see, Costa et al. 1997; van Paradijs et al. 1997;
Frail et al. 1997). Thanks to the observation of the afterglow,
redshifts of many GRBs are available (see, Metzger et al. 1997;
Bloom et al. 1998; Kulkarni et al. 1998; Djorgovski et al. 1998),
favoring the previous proposal that they are cosmological origin.
The largest redshift of GRBs is found to be 9.4 up to date
(Cucchiara et al. 2011).

Statistically, GRBs can be divided into two classes, long-soft
bursts and short-hard bursts (Kouveliotou et al. 1993; Fishman \&
Meegan 1995; Qin et al. 2000). It has been generally believed that
many long-soft bursts are the event associated with the massive star
collapsars whilst many short-hard bursts are produced in the event
of binary neutron star or neutron star-black hole mergers (e.g.,
Eichler et al. 1989; MacFadyen \& Woosley 1999; Paczynski 1986,
1998; Woosley 1993). The most important achievement after the
successful launching of the Swift satellite is that the two
progenitor proposal for GRBs is favored by a large number of
evidence. It was reported that short bursts were found in regions
with lower star-formation rates, and no evidence of supernovae to
accompany them was detected (Barthelmy et al. 2005; Berger et al.
2005; Hjorth et al. 2005). Meanwhile, long bursts were found to be
originated from star-forming regions in galaxies (Fruchter et al.
2006), and in some of these events, supernovae were detected to
accompany the bursts (Hjorth et al. 2003; Stanek et al. 2003).

Several statistical relations between different parameters of GRBs
have been established. Among them, two are very important since they
might be associated with the origin of the explosion energy of GRBs
or the mechanism of the GRB event. One is the $E_{p}-E_{iso}$
relation (Amati et al. 2002) and the other is the $L_{p}-E_{p}$
relation (Yonetoku et al. 2004). These relations were observed from
long bursts, but as Tsutsui et al. (2012) reported, they hold for
short bursts as well (generally, they are 1/10 and 1/100 dimmer than
the ones of long GRBs, respectively). The two relations have obvious
different behaviors when short and long GRBs are separately
considered. It was found that while long bursts form a tight
relation between $E_{p}$ and $E_{iso}$, short GRBs do not follow the
same relation; in fact, short GRBs are likely to follow another
comparable relation (Amati 2006, 2010; Ghirlanda et al. 2009;
Tsutsui et al. 2012; Zhang et al. 2012). Quite the contrary, long
and short bursts were found to follow alomost the same $L_{p}-E_{p}$
relation (Ghirlanda et al. 2009; Tsutsui et al. 2012; Zhang et al.
2012a; Zhang et al. 2012b). This suggests that the two kinds of
burst might differ in the explosion energy (that is likely to give
rise to $E_{iso}$) which must be closely associated with their
progenitors, whilst they might share the same mechanism of radiation
(that might affect $L_{p}$ and $E_{p}$) such as internal or external
shocks. Or, to plainly interpret, the fact that long and short GRBs
do not follow the same $E_{p}-E_{iso}$ relation is likely due to the
difference of their progenitors; the fact that long and short GRBs
do follow the same $L_{p}-E_{p}$ relation might be due to that their
radiation mechanisms are intrinsically the same. The scenario that
long and short bursts share almost the same radiation mechanism but
correspond to different progenitors is a common consensus. It was
favored recently by the investigation of time resolve behaviors of
Fermi bursts (Ghirlanda et al. 2011a, 2011b).

For the radiation mechanism shared by both long and short GRBs, why
and how it gives rise to a certain power law relation between
$L_{p}$ and $E_{p}$ as shown by current GRB data (see Yonetoku et
al. 2004)? What parameters cause this relation? This motivates our
analysis below.

In section 2, we study the kinetic and dynamic effects in terms of
statistics and discuss the implication of the dynamic index in some
typical cases. In section 3, we apply current GRB data to provide
constrains to the dynamic index. Three important issues are
discussed in sections 4-6, and conclusions are presented in section
7.

\section{Kinetic and dynamic effects in terms of statistics}
It is generally believed that radiation of GRBs is generated by
internal or external shocks from the ejecta that moving away from
the central region of the burst with relativistical speeds (Rees \&
Meszaros 1992, 1994; Meszaros \& Rees 1993, 1994; Katz 1994;
Paczynski \& Xu 1994; Sari et al. 1996). In this situation, the
radiation observed must be highly influenced by the relativistic
effect due to the very large speed of motion. In fact, there are two
modes of analyzing the radiation of these objects. One is to
directly consider the radiation in the observer rest frame (or, more
precisely, in the cosmological rest frame), which we call as mode A.
The other is first to consider the radiation in the ejecta rest
frame and then consider how this original radiation is changed by
the motion of the ejecta to that is observed, which we call as mode
B. With mode A, one can display how the light curve as well as the
spectrum of different radiation mechanisms evolve, where the kinetic
and dynamic effects are not separately considered (see, e.g., Granot
\& Sari 2002). When applying mode A, one should know which
particular mechanisms are involved. With mode B, one can clearly
tell how the kinetic effect works, regardless the details of the
radiation mechanisms. For example, with mode B, the kinetic effect
can be revealed in details even though one only knows a certain form
of emission such as the Band function emission form that the ejecta
rest frame radiation bears. This attempt succeeds in the
investigation of the curvature effect, where characteristics of the
kinetic effect arising from an expanding fireball shell can be well
displayed without knowing the details of the ejecta rest frame
radiation (see, e.g., Qin et al. 2004; Qin \& Lu 2005; Qin 2009).
Here, we try to analyze the implication of the $L_p - E_p$ relation
with mode B.

\subsection{The role of the kinetic effect}
Consider the radiation from an ejecta moving with a Lorentz factor
$\Gamma$. Let us assign
\begin{equation}
D\equiv\frac{1}{\Gamma(1-\beta cos\theta)}
\end{equation}
to represent the Doppler effect factor, where $\theta$ is the angle
between the motion direction of the ejecta and the line of sight. In
the theory of special relativity, the observed specific intensity
$I_{\nu}$ ($erg~ s^{-1} cm^{-2} str^{-1} Hz^{-1}$) is related to the
rest frame specific intensity $I_{\nu,0}$ by
$I_{\nu}=D^{3}I_{\nu,0}$, indicating that the Doppler effect can
boost the intensity to a much larger value. (Note that,
$I_{\nu}/\nu^3$ is a Lorentz invariant which keeps the same form
under the transformation between different inertial frames, and
relation $I_{\nu}/\nu^3=I_{\nu,0}/\nu_{0}^3$ leads to
$I_{\nu}=D^{3}I_{\nu,0}$ when the Doppler effect is applied.)

Generally, the observed peak luminosity of the radiation from the
ejecta, $L_{p}$, is proportional to the ejecta rest frame peak
luminosity, $L_{p,0}$, and is a function of the Lorentz factor as
well as the Doppler effect factor:
\begin{equation}
L_{p}=L_{p,0}\Gamma^{\alpha_{G}}D^{\alpha_{D}},
\end{equation}
where, indexes $\alpha_{G}$ and $\alpha_{D}$ depend on the
situation, e.g., the size of emitting material and the viewing
angle.

Let the cosmological rest frame $\nu f_{\nu}$ spectrum peak energy
be $E_{p,r}$ and the ejecta rest frame peak energy be $E_{p,0}$.
According to the Doppler effect we get
\begin{equation}
E_{p,r}=E_{p,0}D.
\end{equation}

Combing equations (2) and (3) gives rise to
\begin{equation}
L_{p}=L_{p,0}\Gamma^{\alpha_{G}}(\frac{E_{p,r}}{E_{p,0}})^{\alpha_{D}}.
\end{equation}
Equation (4) describes the pure kinetic effect in the $L_{p}-E_{p}$
relation. It depends on the Lorentz factor, and it would be
applicable so long as indexes $\alpha_{G}$ and $\alpha_{D}$ are
available.

Unlike in the case of mode A, this analysis does not refer to the
details of radiation. Its result does not depend on any particular
radiation mechanisms. Equation (4) can be well applied to the
investigation of an individual GRB behavior as well as the
investigation of the collective behavior of any GRB samples (say,
the investigation of the statistical properties of GRB samples).

\subsection{The role of the dynamic effect}
Actually, the ejecta rest frame peak luminosity $L_{p,0}$ and the
ejecta rest frame peak energy $E_{p,0}$ themselves might rely on the
Lorentz factor of the ejecta if the radiation is produced by shocks.
If so, they can be expressed as $L_{p,0}=L_{p,0}(\Gamma)$ and
$E_{p,0}=E_{p,0}(\Gamma)$ respectively. These relations might depend
on the strength of the shocks and the type of radiation as well as
other physical conditions such as the environment of the ejecta.

In fact, the functions of $L_{p,0}(\Gamma)$ and $E_{p,0}(\Gamma)$
might be complex rather than simple. However, if the dependence of
$L_{p,0}$ or $E_{p,0}$ on $\Gamma$ is certain or stable, in a wide
range of $\Gamma$ concerned, both $L_{p,0}(\Gamma)$ and
$E_{p,0}(\Gamma)$ can generally be expressed as a power law function
of $\Gamma$, especially when they are studied statistically. For a
mechanism that depends on $\Gamma$ within a narrow range but does
not show a dependence on $\Gamma$ in a wide range (such as in the
case of a periodicity function), then the index of the power law
will be zero in terms of statistics. In this situation, the narrow
range dependence will show a scatter of data in the $L_{p,0}-\Gamma$
or $E_{p,0}-\Gamma$ plane when the data are concerned in a wide
range of $\Gamma$. In the case of a mechanism that depends strongly
on $\Gamma$ within a narrow range but depends weakly on $\Gamma$ in
a wide range, the index of the power law will be small, and in this
situation, the narrow range dependence will show a scatter of data
in the power law relation of $L_{p,0}-\Gamma$ or $E_{p,0}-\Gamma$.
Therefore, in terms of statistics, one can always assume power-law
functions of $L_{p,0}(\Gamma)$ and $E_{p,0}(\Gamma)$ in a wide range
of $\Gamma$. These functions might not fit the actual relations of
$L_{p,0}(\Gamma)$ and $E_{p,0}(\Gamma)$ in a narrow range of
$\Gamma$, but they would fit the relations in a wide range, and this
in turn would impose constrains to the dynamic mechanisms if the
power-law indexes are available from statistical analysis.

Let us assume that
\begin{equation}
L_{p,0}=L_{p,00}\Gamma^{\alpha_{L}}
\end{equation}
and
\begin{equation}
E_{p,0}=E_{p,00}\Gamma^{\alpha_{E}}
\end{equation}
be valid in a wide range of $\Gamma$, where $\alpha_{L}$ and
$\alpha_{E}$ are constants, and $L_{p,00}$ and $E_{p,00}$ are free
of $\Gamma$ but they can vary from source to source.

Applying equations (5) and (6), we get from equation (4) that
\begin{equation}
L_{p}=L_{p,00}\Gamma^{\alpha_{G}+\alpha_{L}-\alpha_{D}
\alpha_{E}}(\frac{E_{p,r}}{E_{p,00}})^{\alpha_{D}}.
\end{equation}
This is a general form of the relation that combines the kinetic and
the statistical dynamic effects. According to equations (3) and (6),
$\Gamma$ and $E_{p,r}$ are related by
\begin{equation}
E_{p,r}=E_{p,00}\Gamma^{\alpha_{E}}D.
\end{equation}
Once the moving direction is known (say, when $\theta$ is provided),
$\Gamma$ would merely be  a function of $E_{p,r}$, and this would
make equation (7) applicable to explore the $L_{p}-E_{p}$ relation.

Generally, GRBs are believed to be the radiation arising from the
ejecta that moves nearly close to the line of sight with a
relativistic speed. Let us consider the situation that the ejecta
moves with $\Gamma \gg 1$, directly towards the observer. In this
case, one finds $D=2\Gamma$. We get from equation (8) that
\begin{equation}
E_{p,r}=2E_{p,00}\Gamma^{1+\alpha_{E}}.
\end{equation}
Inserting it into equation (7) yields
\begin{equation}
L_{p}=2^{\alpha_{D}}L_{p,00}(\frac{E_{p,r}}{2E_{p,00}})^{\alpha_{kin}+\alpha_{dyn}},
\end{equation}
with
\begin{equation}
\alpha_{kin}\equiv\alpha_{G}+\alpha_{D}
\end{equation}
and
\begin{equation}
\alpha_{dyn}\equiv\frac{\alpha_{L}-\alpha_{kin}\alpha_{E}}{1+\alpha_{E}}.
\end{equation}
This is the relation that combines the kinetic and the statistical
dynamic effects under the condition that the ejecta moves with
$\Gamma \gg 1$, directly towards the observer. Revealed in equation
(10), while index $\alpha_{kin}$ describes the pure kinetic effect,
index $\alpha_{dyn}$ corresponds to the statistical dynamic effect.
We call $\alpha_{kin}$ and $\alpha_{dyn}$ as the kinetic and dynamic
indexes in the $L_{p}-E_{p}$ relation, respectively. The value of
$\alpha_{dyn}$ measured from statistical analysis of observational
data would strongly constrain the dynamic mechanisms that are
involved.

\subsection{Implication of cases with typical dynamic indexes}
As the components of the dynamic index, $\alpha_{L}$ reflects how
the radiation strength of the shock in the ejecta rest frame is
connected to the Lorentz factor of the ejecta and $\alpha_{E}$
reflects how the hardness of the radiation spectrum that is produced
by the shock is influenced by the Lorentz factor. Due to the vast
difference of the physical conditions that the ejecta would
encounter, it might be possible that, in some cases, the effects
associated with $\alpha_{L}$ and $\alpha_{E}$ are similar, while in
other cases, they are entirely different.

As shown in equation (9), if $\alpha_{E}= -1$, then $E_{p,r}$ would
not depend on $\Gamma$. This very special situation is not discussed
below. In the following we assume that $\alpha_{E}\neq -1$. As long
as we know, $\alpha_{G}\ge 0$ and $\alpha_{D}\ge 0$ always hold in
the situation concerned, therefore we also assume $\alpha_{kin}\ge
0$ in the following analysis.

It has been generally believed that, the emission of GRBs comes from
internal or external shocks which are produced by the event that an
inner ejecta catches up the outer ejecta or a fast moving ejecta
hits the external medium. Theoretically, we refer the so-called
``ejecta rest frame'' as the frame relative to which there is no
particular collective motion of all the radiation seeds (such as
chaotic moving electrons). For the sake of simplicity, here we
regard the final speed, relative to the cosmological rest frame, of
the ejecta after the shock as the speed of the so-called ``ejecta
rest frame''.

The shocks involved might belong to various kinds, depending on the
actual physical conditions. We pay our attention to the following
two typical types of shock.

One is the kind of shock that is associated with
\begin{equation}
\alpha_{L}>0 \qquad \qquad and \qquad \qquad \alpha_{E}\geq0,
\end{equation}
which we call an external-like shock. This kind of shock can be a
real external shock: a fast moving ejecta hits or sweeps external
medium and that gives rise to shocks. For the same kind of external
medium and the same rest mass of the ejecta, the larger the $\Gamma$
value, the stronger the shock and possibly the harder of the
intrinsic radiation spectrum. This kind of shock can also be an
inner ejecta dominant internal shock: the rest mass of the outer
ejecta is much smaller than that of the inner ejecta, and then the
final value of $\Gamma$ would mainly depend on the motion of the
inner ejecta. Under this condition, for the same kind of the outer
ejecta and the same rest mass of the inner ejecta, the larger the
final value of $\Gamma$, the stronger the shocks and possibly the
harder the intrinsic radiation spectrum.

\begin{deluxetable}{cccc}
\tabletypesize{\scriptsize} \tablecaption{Cases associated with
typical dynamic indexes} \tablewidth{0pt} \tablehead{ \colhead{Case}
& \colhead{($\alpha_{E}$)} & \colhead{($\alpha_{L}$)} &
\colhead{($\alpha_{dyn}$)} } \startdata
1a & $\alpha_{E}>-1$ & $\alpha_{L}>\alpha_{kin}\alpha_{E}$ & $\alpha_{dyn}>0$ \\
1b & $\alpha_{E}<-1$ & $\alpha_{L}<\alpha_{kin}\alpha_{E}$ & $\alpha_{dyn}>0$ \\
2a & $\alpha_{E}>0$ & $\alpha_{L}=\alpha_{kin}\alpha_{E}$ & $\alpha_{dyn}=0$ \\
2b & $\alpha_{E}<0$ & $\alpha_{L}=\alpha_{kin}\alpha_{E}$ & $\alpha_{dyn}=0$ \\
3a & $\alpha_{E}>-1$ & $\alpha_{L}>-\alpha_{kin}$ & $-\alpha_{kin}<\alpha_{dyn}<0$ \\
3b & $\alpha_{E}<-1$ & $\alpha_{L}<-\alpha_{kin}$ & $-\alpha_{kin}<\alpha_{dyn}<0$ \\
4a & $\alpha_{E}>0$ & $\alpha_{L}=-\alpha_{kin}$ & $\alpha_{dyn}=-\alpha_{kin}$ \\
4b & $\alpha_{E}<0$ & $\alpha_{L}=-\alpha_{kin}$ & $\alpha_{dyn}=-\alpha_{kin}$ \\
5a & $\alpha_{E}>-1$ & $\alpha_{L}<-\alpha_{kin}$ & $\alpha_{dyn}<-\alpha_{kin}$ \\
5b & $\alpha_{E}<-1$ & $\alpha_{L}>-\alpha_{kin}$ & $\alpha_{dyn}<-\alpha_{kin}$ \\
\enddata
\end{deluxetable}

The other is the kind of shock that is associated with
\begin{equation}
\alpha_{L}<0 \qquad \qquad and \qquad \qquad \alpha_{E}\leq0,
\end{equation}
which we call an internal-like shock. This kind of shock can be an
outer ejecta dominant internal shock, where the rest mass of the
outer ejecta is assumed to be much larger than that of the inner
ejecta. For the same kind of the inner ejecta (say, when the Lorentz
factors and rest masses of all the inner ejecta concerned are almost
the same) and the same rest mass of the outer ejecta, it is
expectable that a larger final speed of the ejecta would produce a
weaker internal shock and possibly a softer intrinsic radiation
spectrum.

Listed in Table 1 are some cases associated with typical values of
the dynamic index.

Interpretations of these cases are simple. For example, in cases 1a
and 1b, the effect from the dynamic mechanism boosts up that from
the kinetic mechanism. The correlation between $L_p$ and $E_p$ would
become stronger than that in the pure kinetic effect. In case 1a, if
$\alpha_{L}<0$, then condition (14) will be satisfied, and the shock
would be an internal-like shock; if $\alpha_{E}\geq0$, then
condition (13) will be satisfied, and the shock would be an
external-like shock. In case 1b, condition (14) is always satisfied,
and the shock would be an internal-like shock. Therefore, both
internal-like and external-like shocks can give rise to the
situation of $\alpha_{dyn}>0$, so long as the above conditions are
satisfied.

\section{Constrains to the dynamic index by current GRB data}
Spectral forms of GRBs vary significantly. While many of them can be
well fitted by the Band function form (Band et al. 1993), some of
them might be accounted for by other forms such as the
bremsstrahlung, Comptonized and synchrotron radiations (Schaefer et
al. 1994). Even for those with an obvious Band function form, both
the low and high energy indexes vary from source to source (see,
e.g., Preece et al. 2000). As the kinetic effect does not affect the
spectral form, it is accordingly probably that emissions of GRBs are
produced by various radiation processes. In this way, it is hard to
interpret their statistical properties based on a single radiation
mechanism.

However, with the statistical assumptions proposed above (see
equations 5 and 6), statistical properties of the radiation
mechanisms will be able to figure out even though the actual
mechanisms themselves are unaware. This in turn can constrain the
mechanisms and possibly provide glues to reveal them. Here, with the
$L_p - E_p$ relation derived above, we try to draw out some
statistical properties of the dynamic mechanism from current GRB
data.

Many authors have explored the $L_p - E_p$ relationship from various
GRB samples. The relation of $L_p \propto E_{p,r}^\nu$ was confirmed
by different groups of researchers in their statistical
investigations. The index, $\nu$, was found to range from 1.4 to 2.0
(see Yonetoku et al. 2004, 2010; Ghirlanda et al. 2005; Wang et al.
2011; Zhang et al. 2012b). Compared with equation (10), we find
that, in our model, the dynamic index in the $L_{p}-E_{p}$ relation
is confined within $\alpha_{dyn} = (1.4-\alpha_{kin},
2.0-\alpha_{kin})$ by current samples.

As long as we know, in common cases of jets, $\alpha_{kin}\ge 2$
always holds (see, e.g., the analysis below). The constrain of
$\alpha_{dyn} = (1.4-\alpha_{kin}, 2.0-\alpha_{kin})$ suggests that,
in terms of statistics, the dynamic effect of GRBs does not boost up
the kinetic effect, but instead, it partially cancels the latter.
Or, the dynamic effect alone tends to make an anti-correlation
between $L_p$ and $E_p$, which is entirely different from what the
kinetic effect does. In addition, the data reveal that, the kinetic
effect would be much larger than the dynamic effect.

In the following, we show how the dynamic index is constrained by
data in a typical case of jets.

The effect of special relativity on the observed flux of jets in
AGNs and GRBs have been discussed by various authors (e.g., Rybicki
\& Lightman 1979; Lind \& Blandford 1985; Atoyan \& Aharonian 1997;
Sikora et al. 1997; Granot et al. 1999; Woods \& Loeb 1999). Known
as the so-called Doppler boosting effect, the ratio of observed flux
density $S_{obs}$ ($erg~ s^{-1} cm^{-2} Hz^{-1}$) to emitted flux
density $S_0$ from an optically thin, isotropically emitting jet is
$S_{obs}=D^{k-\alpha}S_0$, where $\alpha$ is the spectral index of
the emission ($S_{\nu} \propto \nu^{-\alpha}$), and $k$ is a
parameter that accounts for the geometry of the ejecta, with $k = 2$
for a continuous jet and $k=3$ for a jet of moving discrete radio
clouds (see, e.g., Mirabel \& Rodriguez 1999). Since the luminosity
concerned in this paper is that covers the entire energy range, to
consider the ejecta rest frame peak luminosity, $L_{p,0}$, one
should ignore the effect arising from the Doppler shifting of the
spectrum of the source. Therefore, we prefer to use
$S_{\nu}=D^{k}S_{\nu_{0},0}$ instead of $S_{obs}=D^{k-\alpha}S_0$,
where $\nu$ and $\nu_{0}$ are related by the Doppler effect. For
luminosities corresponding to the sum of the flux over the entire
energy range, $L_{p}=L_{p,0}D^{k}$ is applicable in the two
situations of jets concerned above. In such cases, one finds
$\alpha_{G}=0$ and $\alpha_{D}=2$ for a continuous jet, and
$\alpha_{G}=0$ and $\alpha_{D}=3$ for a jet of moving discrete radio
clouds. They correspond to $\alpha_{kin}=2$ and $\alpha_{kin}=3$
respectively.

Let us consider the case of jets with moving discrete radio clouds.
In this situation, $\alpha_{kin}=3$. Thus, the constrain of the
dynamic index becomes $\alpha_{dyn} = (-1.6, -1)$, and equation (12)
can be expressed as
\begin{equation}
\alpha_{dyn}=\frac{\alpha_{L}-3\alpha_{E}}{1+\alpha_{E}},
\end{equation}
and the indexes in Table 1 can be presented in a more plain way (see
Table 2).

\begin{deluxetable}{cccc}
\tabletypesize{\scriptsize} \tablecaption{Dynamic indexes in Table 1
in the case of jets with moving discrete radio clouds}
\tablewidth{0pt} \tablehead{ \colhead{Case} &
\colhead{($\alpha_{E}$)} & \colhead{($\alpha_{L}$)} &
\colhead{($\alpha_{dyn}$)} } \startdata
1a & $\alpha_{E}>-1$ & $\alpha_{L}>3\alpha_{E}$ & $\alpha_{dyn}>0$ \\
1b & $\alpha_{E}<-1$ & $\alpha_{L}<3\alpha_{E}$ & $\alpha_{dyn}>0$ \\
2a & $\alpha_{E}>0$ & $\alpha_{L}=3\alpha_{E}$ & $\alpha_{dyn}=0$ \\
2b & $\alpha_{E}<0$ & $\alpha_{L}=3\alpha_{E}$ & $\alpha_{dyn}=0$ \\
3a & $\alpha_{E}>-1$ & $\alpha_{L}>-3$ & $-3<\alpha_{dyn}<0$ \\
3b & $\alpha_{E}<-1$ & $\alpha_{L}<-3$ & $-3<\alpha_{dyn}<0$ \\
4a & $\alpha_{E}>0$ & $\alpha_{L}=-3$ & $\alpha_{dyn}=-3$ \\
4b & $\alpha_{E}<0$ & $\alpha_{L}=-3$ & $\alpha_{dyn}=-3$ \\
5a & $\alpha_{E}>-1$ & $\alpha_{L}<-3$ & $\alpha_{dyn}<-3$ \\
5b & $\alpha_{E}<-1$ & $\alpha_{L}>-3$ & $\alpha_{dyn}<-3$ \\
\enddata
\end{deluxetable}

According to the vast variance of the observational characters and
according to the many differences of the physical conditions that
one can imagine (environments, masses, etc.), it is expectable that
GRB events might arise from different kinds of shock and might have
different radiation mechanisms. Perhaps one can divide GRBs into
several subsets merely according to their types of shock and/or
their radiation mechanisms. We suspect that it is these radiation
mechanisms and shocks and the distributions of these subsets that
gives rise to the observed $L_p - E_p$ relation. Thus, assuming a
single mechanism to account for the $L_p - E_p$ relation of the
whole sample seems unreasonable. However, this should not prevent us
to explore what a mechanism could be expected when it is supposed to
account for the relation. Perhaps this approach can provide
constrains to mechanisms or provide glues to find them.

\begin{figure}[tbp]
\begin{center}
\includegraphics[width=3in,angle=0]{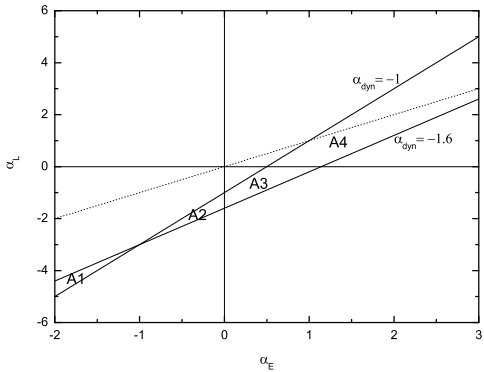}
\end{center}
\caption{Constrains to the ranges of $\alpha_{L}$ and $\alpha_{E}$
in the situation of jets with moving discrete radio clouds by
current GRB data, where the two thick solid lines are the
$\alpha_{dyn} = -1.6$ and $\alpha_{dyn} = -1$ curves, respectively.
The two thin solid lines represent the axes, and the dotted line is
the identity curve.} \label{Fig. 1}
\end{figure}

According to equation (15), once $\alpha_{dyn}$ is known, the
relation between $\alpha_{L}$ and $\alpha_{E}$ will be well
established. Once a range of $\alpha_{dyn}$ is provided, the
available regions of $\alpha_{L}$ and $\alpha_{E}$ will be well
constrained. With $\alpha_{dyn} = (-1.6, -1)$, the constrains to the
ranges of $\alpha_{L}$ and $\alpha_{E}$ in the situation of jets
with moving discrete radio clouds are now available. Displayed in
Fig. 1 are the two marginal curves in the $\alpha_{L} - \alpha_{E}$
plane: the $\alpha_{dyn} = -1.6$ and $\alpha_{dyn} = -1$ curves.
Four areas are confined by these curves and the axes. They are A1,
A2, A3 and A4.

In areas A1 and A2, one finds $\alpha_{L}<0$ and $\alpha_{E}<0$.
This satisfies condition (14), suggesting that parameters confined
within these areas would be associated with internal-like shocks
such as the outer ejecta dominant internal shocks where the rest
mass of the outer ejecta is assumed to be much larger than that of
the inner ejecta.

In area A3, $\alpha_{L}<0$ and $\alpha_{E}>0$. Parameters within
this area correspond to the situation that, a larger value of
$\Gamma$ would give rise to a weaker shock with a harder intrinsic
spectrum. We do not know if any dynamical mechanisms can produce
this kind of shock. If the answer is no, we would prefer to regard
this area merely to be a result contributed by different GRB subsets
which are associated with different mechanisms.

In area A4, $\alpha_{L}>0$ and $\alpha_{E}>0$, satisfying condition
(13). As mentioned above, parameters confined within this area would
be associated with external-like shocks, including external shocks
and inner ejecta dominant internal shocks where the rest mass of the
inner ejecta is assumed to be much larger than that of the outer
ejecta.

Fig. 1 reveals that, in both areas of A1 and A2, where
$\alpha_{L}<0$ and $\alpha_{E}<0$, the relation between the indexes
follows $\alpha_{L}<\alpha_{E}$. This indicates that, in the case of
internal-like shocks, $L_{p,0}$ depends on $\Gamma$ more strongly
than $E_{p,0}$ does. In a large part of area A4 (the upper right
portion of A4, above the identity curve, where the values of the
indexes are relatively larger), the indexes satisfy
$\alpha_{L}>\alpha_{E}$ (here, $\alpha_{L}>0$ and $\alpha_{E}>0$),
suggesting that, in some cases of external-like shocks, $L_{p,0}$
also depends on $\Gamma$ more strongly than $E_{p,0}$ does. However,
in the rest part of area A4 (the lower left portion of A4, below the
identity curve, where the values of the indexes are relatively
smaller), the relation between the indexes is
$\alpha_{L}<\alpha_{E}$, showing that, in some cases of
external-like shocks, $L_{p,0}$ depends on $\Gamma$ more mildly than
$E_{p,0}$ does. This becomes a statistical constrain to the
radiation mechanism.

\section{Discussion 1: what would happen when viewing from the beaming angle}
In the above analysis, we consider only the case that the ejecta
moves directly towards the observer (say, $\theta = 0$). However,
some GRBs might be the events of the emission from the ejecta which
does not move along the direction of the line of sight.

It is general believed that, as the ejecta moves with a relativistic
speed, most part of the detected GRB energy must be the radiation
from the area confined within a very small solid angle relative to
the explosion spot, around the line of sight. This is the so-called
beaming effect. The corresponding angle is the beaming angle which
satisfies $cos\theta = \beta$. Is the emission from the beaming
angle much different from that of $\theta = 0$?

Taking $cos\theta = \beta$, one finds $D=\Gamma$, and then gets from
equation (8) that
\begin{equation}
E_{p,r}=E_{p,00}\Gamma^{1+\alpha_{E}}.
\end{equation}
Inserting it into equation (7) yields
\begin{equation}
L_{p}=L_{p,00}(\frac{E_{p,r}}{E_{p,00}})^{\alpha_{kin}+\alpha_{dyn}},
\end{equation}
where $\alpha_{kin}$ and $\alpha_{dyn}$ are represented by equations
(11) and (12), respectively.

Comparing equations (10) and (17) we find that the $L_{p}-E_{p}$
relation in the case of $cos\theta = \beta$ is almost the same as
that in the case of $\theta = 0$. The index is exactly the same,
while the coefficient is slightly different. Therefore, the result
of the discussion of the dynamic indexes presented above holds in
the case of the beaming angle.

\section{Discussion 2: constrains to the dynamic index in two general situations}
Let us consider two models of jets, which might be more close to
real situations.

\subsection{In the case of steady-state jets with an open
angle $\theta_j \sim 1/\Gamma_j$}
Sikora et al. (1997) showed that, in the case of steady-state jets
with an open angle $\theta_j \sim 1/\Gamma_j$, emitting
isotropically in the jet frame, when viewing from the angle of
$\theta_{obs} \sim 1/\Gamma_j$, the relation between the observed
luminosity and the rest frame luminosity may be $L_{obs}\cong
2\Gamma_j^{2}L_{em}$, where $\Gamma_j$ is the bulk Lorentz factor of
the jets.

Viewing angle $\theta = 1/\Gamma$ is equivalent to $cos\theta =
\beta$, which is the well-known beaming angle. Therefore, for
$\theta_{obs} \sim 1/\Gamma_j$, one finds $D \sim \Gamma_j$. Then
equation (3) gives rise to $E_{p,r}\sim E_{p,0} \Gamma_j$. Applying
equations (5) and (6), equation (10) should be replaced by
\begin{equation}
L_{p}\cong
2L_{p,00}(\frac{E_{p,r}}{E_{p,00}})^{\alpha_{kin}+\alpha_{dyn}},
\end{equation}
with $\alpha_{kin}=2$ and
$\alpha_{dyn}=(\alpha_{L}-2\alpha_{E})/(1+\alpha_{E})$. In this
situation, the constrain of the dynamic index by current GRB data
becomes $\alpha_{dyn} = (-0.6, 0)$.

\begin{figure}[tbp]
\begin{center}
\includegraphics[width=3in,angle=0]{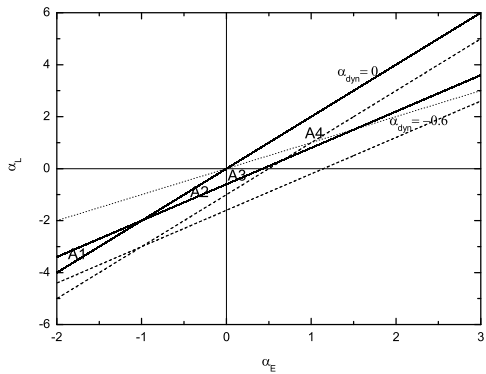}
\end{center}
\caption{Constrains to the ranges of $\alpha_{L}$ and $\alpha_{E}$
in the situation of steady-state jets with an open angle $\theta_j
\sim 1/\Gamma_j$, viewing from the angle of $\theta_{obs} \sim
1/\Gamma_j$, by current GRB data, where the two thick solid lines
stand for the $\alpha_{dyn} = -0.6$ and $\alpha_{dyn} = 0$ curves,
respectively. For the sake of comparison, the two marginal curves in
Fig. 1 are also presented (the dashed lines). Other symbols are the
same as in Fig. 1.} \label{Fig. 1}
\end{figure}

Shown in Fig. 2 are the corresponding marginal curves in the
$\alpha_{L} - \alpha_{E}$ plane: the $\alpha_{dyn} = -0.6$ and
$\alpha_{dyn} = 0$ curves. As in the case of jets with moving
discrete radio clouds, four areas, A1, A2, A3 and A4, are confined
by these curves and the axes. Interpretations of these areas remain
the same as those in Fig. 1. Fig. 2 reveals that, the constrains to
the dynamical indexes rely strongly on the model of jets. A model
with a smaller value of $\alpha_{kin}$ would shift the constrain
areas to the domain of larger $\alpha_{L}$.

\subsection{In the case of jets with the moving blobs
radiating within a distance range}
Sikora et al. (1997) also studied the situation of jets with moving
blobs which radiate within a distance range. They assumed that (1)
the blobs are injected into the ``active zone'' every $\Delta
t_{inj}$; (2) all blobs are moving with the same Lorentz factor; (3)
blobs radiate isotropically in their rest frame, each at the same
rate, $L'_{em,1}$; (4) each blob stops to radiate after passing a
given distance range $\Delta r=r$. They also considered the case of
viewing angle of $\theta_{obs} \sim 1/\Gamma_j$. In this situation,
they found the relation of $L_{obs} = L_{em}D^4$, where
$L_{em}=L'_{em}=N_{obs}L'_{em,1}$, and $N_{obs}$ is the number of
blobs counted by a distant observer within his observational time
interval $\Delta t_{obs}$.

Repeating the above analysis we get
\begin{equation}
L_{p} =
L_{p,00}(\frac{E_{p,r}}{E_{p,00}})^{\alpha_{kin}+\alpha_{dyn}},
\end{equation}
with $\alpha_{kin}=4$ and
$\alpha_{dyn}=(\alpha_{L}-4\alpha_{E})/(1+\alpha_{E})$. In this
situation, the constrain of the dynamic index by current GRB data
becomes $\alpha_{dyn} = (-2.6, -2)$.

\begin{figure}[tbp]
\begin{center}
\includegraphics[width=3in,angle=0]{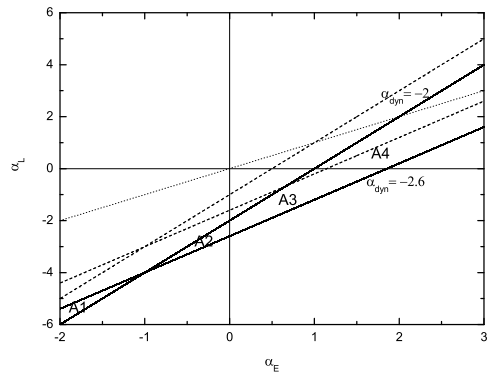}
\end{center}
\caption{Constrains to the ranges of $\alpha_{L}$ and $\alpha_{E}$
in the situation of jets with the moving blobs radiating within a
distance range, viewing from the angle of $\theta_{obs} \sim
1/\Gamma_j$, by current GRB data, where the two thick solid lines
stand for the $\alpha_{dyn} = -2$ and $\alpha_{dyn} = -2.6$ curves,
respectively. The dashed lines represent the two marginal curves in
Fig. 1. Other symbols are the same as in Fig. 1.} \label{Fig. 1}
\end{figure}

Contrary to that hinted by Fig. 2, Fig. 3 shows that, with current
GRB data, a model with a larger value of $\alpha_{kin}$ would shift
the constrain areas to the domain of smaller $\alpha_{L}$.

\section{Discussion 3: normalization of the relation}
It should be pointed out that, the $L_p - E_p$ relation has two
parameters, i.e. index and normalization. In the analysis above, we
only discuss the slope index in view of kinetic and dynamic effects.
Here, let us discuss how the kinetic and dynamic effects influence
the normalization of the relation.

Comparing equations (10), (17), (18) and (19), we find that the
normalization differs from model to model. It is proportional to
$L_{p,00}$ and $E_{p,00}^{-\alpha_{kin}-\alpha_{dyn}}$. The latter
strongly depends on both kinetic and dynamic indexes. The relation
between the normalization and $E_{p,00}$ varies for different
situations of jets. To interpret the normalization, one should tell
which model of jets is concerned.

Let us consider the simplest model of jets: jets with moving
discrete radio clouds which moves directly towards the observer. In
this situation, equation (10) becomes
\begin{equation}
L_{p}=8L_{p,00}(\frac{E_{p,r}}{2E_{p,00}})^{3+\alpha_{dyn}}.
\end{equation}
The normalization is
$2^{-\alpha_{dyn}}L_{p,00}E_{p,00}^{-3-\alpha_{dyn}}$.

According to the above analysis we know that, in this situation,
$L_{p,00}$, $E_{p,00}$ and $\alpha_{dyn}$ are products of the
dynamic effect, while index $3$ is the result of the kinetic effect.
For the proposed model, while the kinetic effect is certain, the
dynamic effect is not constrained. Let us assume that all sources
concerned obey relation (20), where $\alpha_{dyn}$ is certain while
$L_{p,00}$ and $E_{p,00}$ are allowed to vary from source to source.
In this way, the dispersion of the normalization will only come from
that of $L_{p,00}$ and $E_{p,00}$. Since $L_{p,00}$ and $E_{p,00}$
are the two basic parameters of the normalization, the dynamic
effect is important to evaluate the quantity.

To our surprise, the kinetic effect influences the normalization as
well. For example, if $E_{p,00}$ is $10$ times smaller than that
expected, then the kinetic effect alone will make the normalization
$1000$ times smaller than the expected one. Therefore, to study the
normalization, the kinetic effect should not be ignored. In fact,
the kinetic effect affects the normalization in two aspects: the
kinetic index, and the formula itself (compare equations 10, 17, 18
and 19).

As an example of discussion, let us study the normalization of the
$L_p - E_p$ relation obtained by Yonetoku et al. (2004):
$(2.34^{+2.29}_{-1.76})\times 10^{-5}$, where $L_p$ is in units of
$10^{52}~ erg~ s^{-1}$ and $E_p$ is in units of $keV$. In their
work, the index of $E_p$ is $2$.

Here, we adopt the model of jets considered in this section (i.e.,
jets with moving discrete radio clouds which moves directly towards
the observer). In this way, equation (20) is applicable. For this
equation, the data suggest $\alpha_{dyn}=-1$. Therefore, the
normalization of equation (20) becomes $2L_{p,00}E_{p,00}^{-2}$,
where $L_{p,00}$ is in units of $10^{52}~ erg~ s^{-1}$ and
$E_{p,00}$ in units of $keV$.

Hence we approximately have $2L_{p,00}E_{p,00}^{-2}=2.34\times
10^{-5}$. If we assume that $E_{p,00}=1keV$, then we will get the
rest frame luminosity $L_{p,00} \sim 10^{47}~ erg~ s^{-1}$. If we
assume that $E_{p,00}=0.1keV$, then we will get $L_{p,00} \sim
10^{49}~ erg~ s^{-1}$. If we believe that $E_{p,00}$ is around the
range of $0.1keV$ to $1keV$ and the situation concerned above can
approximately represent the real situation of jets, then we will
find that the rest frame luminosity is around the range of $10^{47}~
erg~ s^{-1}$ to $10^{49}~ erg~ s^{-1}$.

Of course, this interpretation is not conclusive since we do not
know the real range of $E_{p,00}$, nor we know the real situation of
jets. However, the above discussion suggests that, with our method,
one can estimate the rest frame luminosity of jets so long as the
range of the rest frame peak energy and the situation of jets are
somehow known (perhaps from other independent investigations).

We suggest that, in a detailed discussion of the normalization of
the $L_p - E_p$ relation in the near future, one should define and
collect a neat subset of sources, assuming that all the sources
belong to a single model. In this way, the formulas applied would be
certain, and hence the index of the kinetic effect will be
available, and this will make the form of the normalization more
certain. (In fact, for such a subset, one might find a tighter
correlation between the two quantities and that the index will be
better established). In this case, it might be expectable that both
$L_{p,00}$ and $E_{p,00}$ would not vary violently from source to
source, and this will certainly be helpful for a quantitative study
of the normalization.

\section{Conclusions}
In this paper, we investigate the underlying mechanisms of the $L_p
- E_p$ relation by separately considering the corresponding kinetic
and dynamic mechanisms. In this way, one can tell how much the
kinetic or dynamic mechanism contributes to the index of the
relationship once a set of GRB data are available.

Based on the analysis above, we reach several conclusions. The first
relies only on theoretical analysis and is quite robust.

a) The kinetic effect alone would give rise to an obvious
correlation between $L_p$ and $E_p$, with the index of the
correlation being as large as 2 to 4, depending on the situation
concerned. This conclusion comes from the effect of the theory of
special relativity, which is entirely independent of observational
data.

Others rely strongly on observation. The following are derived from
current GRB data.

b) The dynamic effect alone tends to make an anti-correlation
between $L_p$ and $E_p$, which partially cancels the kinetic effect.

c) In terms of statistics, the dynamic effect is much smaller than
the kinetic effect.

d) In the case of jets with moving discrete radio clouds when they
move directly towards the observer, the index of the dynamic effect
in the $L_p - E_p$ relation is currently constrained within
$\alpha_{dyn} = (-1.6, -1)$, while in the case of other models of
jets, the constrains are different.

e) Both internal and external shocks can account for the current
data.

From Figs. 2 and 3 one finds that, while the constrain areas (A1,
A2, A3 and A4) would become much narrower in the near future when
much larger samples of data are available, uncertainties of the
dynamic indexes would remain to be large if the model involved is
uncertain. To get a more precise result, perhaps one should discuss
subsets of bursts by assuming that sources of each subset belong to
a single model of jets.

\acknowledgments

This work was supported by the National Natural Science Foundation
of China (No. 11073007) and the Guangzhou technological project (No. 11C62010685).\\


\begin{thebibliography}{99}
\bibitem[\protect\citeauthoryear{}{}]{} Amati, L. 2006, MNRAS, 372, 233
\bibitem[\protect\citeauthoryear{}{}]{} Amati, L. 2010, Journal of the Korean Physical Society, 56, 1603
\bibitem[\protect\citeauthoryear{}{}]{} Amati, L., Frontera, F., Tavani, M., et al. 2002, A\&A, 390, 81
\bibitem[\protect\citeauthoryear{}{}]{} Atoyan, A. M., \& Aharonian F. A. 1997, ApJ, 490, L149
\bibitem[\protect\citeauthoryear{}{}]{} Band, D., Matteson, J., Ford, L., et al. 1993, ApJ, 413, 281
\bibitem[\protect\citeauthoryear{}{}]{} Barthelmy, S. D., Chincarini, G., Burrows, D. N., et al. 2005, Nature, 438, 994
\bibitem[\protect\citeauthoryear{}{}]{} Berger, E., Price, P. A., Cenko, S. B., et al. 2005, Nature, 438, 988
\bibitem[\protect\citeauthoryear{}{}]{} Bloom, J. S., Djorgovski, S., Kulkarni, S. R., \& Frail, D. A. 1998, ApJ, 507, L25
\bibitem[\protect\citeauthoryear{}{}]{} Costa, E., et al. 1997, Nature, 387, 783
\bibitem[\protect\citeauthoryear{}{}]{} Cucchiara, A., Levan, A. J., Fox, D. B., et al. 2011, ApJ, 736, 7
\bibitem[\protect\citeauthoryear{}{}]{} Djorgovski, S. G., et al. 1998, Astrophys. J., 508, L17
\bibitem[\protect\citeauthoryear{}{}]{} Eichler, D., Livio, M., Piran, T., et al. 1989, Nature, 340, 126
\bibitem[\protect\citeauthoryear{}{}]{} Fishman, G. J., Meegan, C. A. 1995, ARA\&A, 33, 415
\bibitem[\protect\citeauthoryear{}{}]{} Frail, D. A., et al. 1997, Nature, 389, 261
\bibitem[\protect\citeauthoryear{}{}]{} Fruchter, A. S., Levan, A. J., Strolger, L., et al. 2006, Nature, 441, 463
\bibitem[\protect\citeauthoryear{}{}]{} Ghirlanda, G., Ghisellini, G., Firmani, C., et al. 2005, MNRAS, 360, L45
\bibitem[\protect\citeauthoryear{}{}]{} Ghirlanda, G., Ghisellini, G., Nava, L., Burlon, D. 2011a, MNRAS, 410, L47
\bibitem[\protect\citeauthoryear{}{}]{} Ghirlanda, G., Ghisellini, G., Nava, L. 2011b, MNRAS, 418, L109
\bibitem[\protect\citeauthoryear{}{}]{} Ghirlanda, G., Nava, L., Ghisellini, G., et al. 2009, A\&A, 496, 585
\bibitem[\protect\citeauthoryear{}{}]{} Granot, J., Piran, T., \& Sari, R. 1999, ApJ, 513, 679
\bibitem[\protect\citeauthoryear{}{}]{} Granot, J., \& Sari, R. 2002, ApJ, 568, 820
\bibitem[\protect\citeauthoryear{}{}]{} Hjorth, J., Sollerman, J., Moller, P., et al. 2003, Nature, 423, 847
\bibitem[\protect\citeauthoryear{}{}]{} Hjorth, J., Watson, D., Fynbo, J. P. U., et al. 2005, Nature, 437, 859
\bibitem[\protect\citeauthoryear{}{}]{} Katz, J. I. 1994, ApJ, 422, 248
\bibitem[\protect\citeauthoryear{}{}]{} Klebesadel, R., Strong, I., \& Olson, R. 1973, ApJ, 182, L85
\bibitem[\protect\citeauthoryear{}{}]{} Kouveliotou, C., Meegan, C. A., Fishman, G. J., et al. 1993, \apj, 413, L101
\bibitem[\protect\citeauthoryear{}{}]{} Kulkarni, S. R., et al. 1998, Nature, 393, 35
\bibitem[\protect\citeauthoryear{}{}]{} Lind, K. R., \& Blandford, R. D. 1985, ApJ, 295, 358
\bibitem[\protect\citeauthoryear{}{}]{} MacFadyen, A. I., Woosley, S. E. 1999, \apj, 524, 262
\bibitem[\protect\citeauthoryear{}{}]{} Metzger, M. R., et al. 1997, Nature, 387, 878
\bibitem[\protect\citeauthoryear{}{}]{} Meszaros, P., \& Rees, M. J. 1993, ApJ, 405, 278
\bibitem[\protect\citeauthoryear{}{}]{} Meszaros, P., \& Rees, M. J. 1994, MNRAS, 269, L41
\bibitem[\protect\citeauthoryear{}{}]{} Mirabel, I. F., \& Rodriguez, L. F. 1999, ARA\&A, 37, 409
\bibitem[\protect\citeauthoryear{}{}]{} Paczynski, B. 1986, \apj, 308, L43
\bibitem[\protect\citeauthoryear{}{}]{} Paczynski, B. 1998, \apj, 494, L45
\bibitem[\protect\citeauthoryear{}{}]{} Paczynski, B., \& Xu, G. 1994, ApJ, 427, 708
\bibitem[\protect\citeauthoryear{}{}]{} Preece, R. D., Briggs, M. S., Mallozzi, R. S., et al. 2000, ApJS, 126, 19
\bibitem[\protect\citeauthoryear{}{}]{} Qin, Y.-P. 2009, ApJ, 691, 811
\bibitem[\protect\citeauthoryear{}{}]{} Qin, Y.-P., \& Lu, R.-J. 2005, MNRAS, 362, 1085
\bibitem[\protect\citeauthoryear{}{}]{} Qin, Y.-P., Xie, G.-.Z, Xue, S.-J., et al. 2000, PASJ, 52, 759
\bibitem[\protect\citeauthoryear{}{}]{} Qin, Y.-P., Zhang, Z.-B., Zhang, F.-W., \& Cui, X.-H. 2004, ApJ, 617, 439
\bibitem[\protect\citeauthoryear{}{}]{} Rees, M. J., \& Meszaros, P. 1992, MNRAS, 258, 41
\bibitem[\protect\citeauthoryear{}{}]{} Rees, M. J., \& Meszaros, P. 1994, ApJ, 430, L93
\bibitem[\protect\citeauthoryear{}{}]{} Rybicki, G. B., \& Lightman, A. P. 1979, Radiative Processes in Astrophysics (New York:Wiley Interscience)
\bibitem[\protect\citeauthoryear{}{}]{} Sari, R., Narayan, R., \& Piran, T. 1996, ApJ, 473, 204
\bibitem[\protect\citeauthoryear{}{}]{} Schaefer, B. E., Teegaeden, B. J., Fantasia, S. F., et al. 1994, ApJS, 92, 285
\bibitem[\protect\citeauthoryear{}{}]{} Sikora, M., Madejski, G., Moderski, R., Poutanen, J. 1997, ApJ, 484, 108
\bibitem[\protect\citeauthoryear{}{}]{} Stanek, K. Z., Matheson, T., Garnavich, P. M., et al. 2003, \apj, 591, L17
\bibitem[\protect\citeauthoryear{}{}]{} Tsutsui, R., Yonetoku, D., Nakamura, T., Takahashi, K., Morihara, Y. 2012, arXiv1208.0429
\bibitem[\protect\citeauthoryear{}{}]{} van Paradijs, J., et al. 1997, Nature, 386, 686
\bibitem[\protect\citeauthoryear{}{}]{} Wang, Fa-Yin, Qi, Shi; Dai, Zi-Gao 2011, MNRAS, 415, 3423
\bibitem[\protect\citeauthoryear{}{}]{} Woods, E., \& Loeb, A. 1999, ApJ, 523, 187
\bibitem[\protect\citeauthoryear{}{}]{} Woosley, S. E. 1993, \apj, 405, 273
\bibitem[\protect\citeauthoryear{}{}]{} Yonetoku, D., Murakami, T., Nakamura, T., et al. 2004, ApJ, 609, 935
\bibitem[\protect\citeauthoryear{}{}]{} Yonetoku, D., Murakami, T., Tsutsui, R., Nakamura, T., Morihara, Y., Takahashi, K. 2010, PASJ, 62, 1495
\bibitem[\protect\citeauthoryear{}{}]{} Zhang, F.-W., Shao, L., Yan, J.-Z., Wei, D.-M. 2012a, ApJ, 750, 88
\bibitem[\protect\citeauthoryear{}{}]{} Zhang, Z. B., Chen, D. Y., Huang, Y. F. 2012b, ApJ, 755, 55
\end{thebibliography}
\end{document}